\documentclass[prd,a4paper,showpacs,
nofootinbib
]{revtex4}
\usepackage{graphicx}
\usepackage{epsfig}
\def\eq#1{{Eq.~(\ref{#1})}}

\newcommand{\be}{\begin{equation}}
\newcommand{\ee}{\end{equation}}
\newcommand{\bea}{\begin{eqnarray}}
\newcommand{\eea}{\end{eqnarray}}

\newcommand{\ms}{\Delta m^2_{\odot}}

\newcommand{\sss}{\sin^2 \theta_{\odot}}

\newcommand{\kl}{\mbox{KamLAND~}}
\newcommand{\beq}{\begin{equation}}
\newcommand{\eeq}{\end{equation}}

\def\anue{{\bar\nu_e}}

\newcommand{\dm}{\mbox{$\Delta m^2$~}}

\def\ltap{\ \raisebox{-.4ex}{\rlap{$\sim$}} \raisebox{.4ex}{$<$}\ }

\newcommand{\deltasol}{\mbox{$ \Delta m^2_{\odot}$}}
\def\ltap{\ \raisebox{-.4ex}{\rlap{$\sim$}} \raisebox{.4ex}{$<$}\ }

\newcommand{\dma}{\mbox{$\Delta m^2_{\rm A}$ }}
\begin{document}

\begin{flushright}
SISSA 77/2003/EP\\
SINP-TNP/03-34\\
hep-ph/0309236
\end{flushright}

\title{On the Measurement of Solar Neutrino Oscillation Parameters 
with KamLAND}

\author{Abhijit Bandyopadhyay$^{1}$}
\author{Sandhya Choubey$^{2,3}$}
\author{Srubabati Goswami$^4$}
\author{S.T. Petcov$^{3,2,5}$}
\affiliation{$^1$Saha Institute of Nuclear Physics,1/AF, Bidhannagar,
Calcutta 700 064, INDIA}
\affiliation{$^2$INFN, Sezione di Trieste, Trieste, Italy}
\affiliation{$^3$Scuola Internazionale Superiore di Studi Avanzati, 
I-34014 Trieste, Italy}
\affiliation{$^4$Harish-Chandra Research Institute, Chhatnag Road, Jhusi,
Allahabad  211 019, INDIA}
\affiliation{$^5$Institute of Nuclear Research and
Nuclear Energy, Bulgarian Academy of Sciences, 1784 Sofia, Bulgaria}

\begin{abstract}
\vskip 1cm
A new reactor power plant Shika-2, with a power 
of approximately  4 GW and at a distance of about 88 km 
from the KamLAND detector is scheduled to start 
operating in March 2006.
We study the impact of the $\anue$ flux from this 
reactor on the sensitivity 
of the KamLAND experiment to the solar neutrino oscillation 
parameters. 
We present results on
prospective determination of  
$\Delta m^2_\odot$ and $\sin^2\theta_\odot$ using the combined data from 
KamLAND and the solar neutrino experiments,
including the effect of the Shika-2  
contribution to the KamLAND signal and 
the latest data from the 
salt enriched phase of the SNO experiment.   
We find that contrary to the expectations, the addition of the 
Shika-2 reactor flux
does not improve the $\sin^2\theta_\odot$ sensitivity of KamLAND, while 
the ambiguity in $\Delta m^2_\odot$ measurement may even 
increase, as a result of the averaging effect between 
Kashiwazaki and the Shika-2 reactor contributions to the 
KamLAND signal. 

\end{abstract}



\maketitle


\section{Introduction}
\label{section:introduction}

   There has been a significant progress in 
the studies of neutrino 
oscillations in the last two years. 
The evidences
for solar neutrino oscillations/transitions,
obtained in the solar neutrino experiments
Homestake, Kamiokande, SAGE, GALLEX/GNO,
Super-Kamiokande (SK) \cite{Cl98,SKsol}
were  reinforced 
by the first data of the SNO experiment 
\cite{SNO1}
on the charged current (CC) 
reaction induced by solar neutrinos,
$\nu_e + D \rightarrow e^{-} + p + p$. 
When combined with the data 
from the Super-Kamiokande
experiment \cite{SKsol}, 
the SNO results 
clearly demonstrate the 
presence of $\nu_{\mu,\tau}$ 
and/or $\bar{\nu}_{\mu,\tau}$
component
in the flux of solar neutrinos reaching the Earth.
This compelling evidence for 
oscillations and/or transitions 
of the solar neutrinos          
was further strengthened in 2002
by the SNO results \cite{SNO2},
including the data 
on the neutral current (NC) reaction 
$\nu + D \rightarrow \nu + n + p$
due to solar neutrinos.
In 2002 the first data from the KamLAND 
experiment were published \cite{KamLAND}.
The KamLAND results represent the first 
strong evidence (more then 3$\sigma$ effect) 
for neutrino oscillations 
obtained in an experiment
with terrestrial neutrinos.
Under the  
assumption of CPT-invariance,
the KamLAND data practically 
establish \cite{KamLAND} the
large mixing angle (LMA)
MSW solution as unique solution
of the solar neutrino problem.
All other mechanisms which could cause
transitions of the solar $\nu_e$ 
into $\nu_{\mu,\tau}$
and/or $\bar{\nu}_{\mu,\tau}$
and which were considered as possible
solutions of the solar neutrino
``puzzle'', such as VO, SMA MSW, QVO, LOW
(see, e.g., \cite{STPSchlad97}),
RSFP \cite{RSFP}, FCNC \cite{FCNC}, WEPV and 
LIV \cite{VWEPVLI},
if not completely ruled out,
are constrained by
the KamLAND data to play at most 
a sub-dominant secondary role in the 
physics of the solar neutrino transitions.
This result
brings us, after more than 
30 years of research, 
initiated by the pioneering
works of B. Pontecorvo \cite{Pont4667} and the
experiment of R. Davis et al. \cite{Davis68},
very close to a complete understanding of the 
true cause of the solar neutrino problem.

  The combined two-neutrino oscillation 
analyses of the available
solar neutrino and KamLAND  data
identified two distinct solution sub-regions within 
the LMA solution region~
(see, e.g., \cite{solfit1,solfit2}).
The best fit values of the two-neutrino 
oscillation parameters - the solar neutrino mixing
angle $\theta_{\odot}$ 
and the neutrino mass squared difference
$\deltasol$,
in the two sub-regions - 
low-LMA and 
high-LMA,
were given respectively by \cite{solfit1}:
$\deltasol = 7.2 \times 10^{-5}~{\rm eV^2}$, 
$\sin^2 \theta_\odot = 0.3$ 
$\deltasol = 1.5 \times 10^{-4}~{\rm eV^2}$, 
$\sin^2 \theta_\odot = 0.3$ 
\noindent The low-LMA solution was statistically preferred 
by the data. At 99.73\% C.L. the two solutions merged and 
the allowed ranges were \cite{solfit1}: 
\begin{equation}
\deltasol \cong 
(5.0 - 20.0) \times 10^{-5}~{\rm eV^2},~~~ 
\sin^2 \theta_\odot \cong (0.21 - 0.47)~.
\label{sol90}
\end{equation}
%

\indent Very recently the SNO collaboration published 
the data from the salt phase of the experiment \cite{SNO3}.
Addition of NaCl in the heavy water increases the efficiency of 
capture for the final state neutrons of the NC reaction resulting 
in a gamma ray cascade with a peak around 8 MeV. 
This, apart from increasing the statistics of the NC data allowed 
the SNO collaboration to report the CC and NC total event rates 
with a higher precision and without the assumption of an undistorted 
energy spectrum. 
For the ratio of the CC and NC event rates, in particular,
the collaboration finds: $R_{CC/NC} = 0.306 \pm 0.026 \pm 0.024$.
Adding the statistical and the systematic errors in quadratures
one gets at 99.73\% C.L.: $R_{CC/NC} \leq 0.41$. 
As was shown in \cite{Maris:2002cv},
an upper limit of $R_{CC/NC} < 0.5$ implies a significant
upper limit on $\ms$ smaller than $2\times 10^{-4}~{\rm eV^2}$.
Thus, the new SNO data on 
$R_{CC/NC}$ implies stringent constraints 
on the high-LMA solution.
A
combined analysis 
of the data from the solar neutrino and KamLAND experiments,
including the latest SNO results,
shows \cite{SNO3ADSSS,others} that the high-LMA solution is allowed 
only at 99.73\% C.L. and restricts the 3$\sigma$ allowed 
upper limit of $\ms$ to 
$\ms < 1.7\times 10^{-4}$ eV$^2$.
 
\indent There exists also very strong 
evidences for oscillations
of the atmospheric $\nu_{\mu}$ ($\bar{\nu}_{\mu}$) 
from the observed Zenith angle dependence of 
the multi-GeV $\mu-$like events in the 
SK experiment \cite{SKatm9802}.
The Super-Kamiokande (SK) atmospheric neutrino data
is best described
in terms of dominant $\nu_{\mu} \rightarrow \nu_{\tau}$ 
($\bar{\nu}_{\mu}\rightarrow \bar{\nu}_{\tau}$) oscillations
with  maximal mixing and 
$1.3 \times 10^{-3} \mbox{eV}^2 \ltap
|\dma| \ltap 3.1 \times 10^{-3} \mbox{eV}^2$
($90\%$ C.L.), where we have quoted the 
preliminary results of an improved analysis of the SK atmospheric 
neutrino data, performed recently by the SK collaboration \cite{SKatmo03}.

  After these remarkable developments
in the field of neutrino oscillations,
further progress in the 
studies of neutrino mixing
requires the realization of 
large and challenging program of 
research in neutrino physics (see, e.g., \cite{SPNDM03}).
As one of its main goals we see
higher precision determination of 
the neutrino mass and mixing parameters
which control the solar 
neutrino oscillations, 
$\deltasol$ and $\theta_{\odot}$.
With the future data from SNO on the day-night effect
and the spectrum of $e^-$ from the CC reaction, and 
the future high statistics data from KamLAND, one can look forward,
in particular, to resolving completely 
the low-LMA - high-LMA solution ambiguity, 
and at the same time constraining the solar neutrino mixing angle 
\cite{Maris:2002cv,solfit1,Maris:2000DN,prekl,preklall}.

 
In general, the solar neutrino oscillations
parameters $\ms$ and $\sss$ can be 
measured either in  solar neutrino experiments, 
or in long baseline 
reactor antineutrino experiments. 
The potential of the current and future solar 
neutrino experiments for precision measurements of 
$\ms$ and $\sss$
was studied in detail recently in
\cite{th12,Bahcall:2003ce}. The just published 
high statistics 
NC data from SNO together with the earlier SNO data
constrain 
the mixing angle $\sss$ from above:
the case of maximal mixing is 
excluded by these data at more than 5 s.d.
The data from the proposed 
low energy experiments sensitive to the $pp$ neutrinos
could help constrain $\sss$ from below, 
provided the total error in these experiments 
could be reduced to $\ltap 1\%$ \cite{th12,Bahcall:2003ce}. 
The recent SNO salt data has also disfavoured higher values of 
\dm and the high-LMA region is shown to be allowed 
only at 3$\sigma$ \cite{SNO3ADSSS,others}. 
In \cite{SNO3ADSSS} it is shown that with 1 kTy statistics 
the \kl data can further narrow down the high-LMA allowed 
zone if the true \kl spectrum corresponds to values of \dm 
in the low-LMA region. However, if the observed \kl spectrum 
corresponds to high-LMA solution, the low-LMA/high-LMA ambiguity 
would increase as a result of conflicting trends of solar and \kl 
data.  

Precision measurements of the solar neutrino oscillation parameters 
can be performed in reactor $\bar{\nu}_e$ experiments 
sensitive to the modulation of the antineutrino spectrum 
caused by the $\ms-$driven oscillations in vacuum. The possibility 
to observe such a modulation depends crucially on the chosen baseline.
When the baseline corresponds to $\sin^2(\Delta m^2_\odot L/E) \approx 1$, 
the neutrinos undergo maximum flavor transition and in this case 
the survival probability $P_{ee} \approx 1-\sin^22\theta_\odot$ is 
in its minimum. 
We will use the term Survival Probability MINima (SPMIN) to denote 
such a case.
If the baseline corresponds to 
$\sin^2(\Delta m^2_\odot L/E)\approx   0$, one has $P_{ee} \approx 1$ 
and we will refer to this case as 
Survival Probability MAXima (SPMAX). While for a given reactor 
experiment both conditions depend crucially on $\Delta m^2_\odot$, 
in the case of the SPMAX condition there is no $\theta_\odot-$dependence. 
Thus, while 
$\Delta m^2_\odot$ can be determined with a relatively good precision 
in an experiment sensitive to 
either SPMAX or SPMIN, for achieving a relatively precise
measurement of the mixing angle $\theta_\odot$  
the baseline should be tuned to SPMIN.

 All studies of the potential of \kl to 
measure the solar neutrino oscillation 
parameters \cite{solfit1,prekl,preklall,th12},
have shown that the \kl experiment 
has a remarkable sensitivity to 
$\ms$. Its expected sensitivity to $\sss$ was not found to be 
equally good \cite{th12}. Even with 3 kTy statistics and extremely 
optimistic systematic error of only 3\%, the constraints on $\sss$ were found 
to be not much better than that obtained using the solar neutrino data 
(see \cite{th12} for a detailed discussion). The reason for this 
can be traced to the fact that effectively the baseline relevant in the 
\kl experiment corresponds to SPMAX for both the 
low-LMA and high-LMA solutions.

   In \cite{th12} a 70 km baseline 
reactor $\bar{\nu}_e$ experiment was proposed, 
to measure with high precision the solar
neutrino oscillation parameters in the case of the low-LMA 
solution. If 
high-LMA happens to be the true solution -- a possibility
which is strongly disfavored by the current data -- an 
intermediate baseline reactor experiment with
$L=20\div 30$ km would be needed for the purpose
\cite{Petcov:2001sy,th12hlma}. 
Such an experiment would not only allow measurement of 
$\ms$ and $\sss$ with high precision, 
but could also constrain $\sin^2\theta$  
and $\dma$ \cite{th12hlma}. Under rather
demanding conditions, the same experiment could provide
information on the neutrino mass hierarchy
\cite{Petcov:2001sy,th12hlma}.

  Both proposals for new reactor experiments 
discussed briefly above were made after 
detailed studies of the \kl physics potential
in which prolonged periods 
of data taking were assumed. 
However, there are new upcoming reactor power plants 
relatively close to \kl. Their presence may 
affect the conclusions reached in these studies. 
More concretely, there are plans to build a new reactor, 
Shika-2, very close to the site of the existing rector 
Shika-1, which is located at about 
88 km from KamLAND \cite{shika2}. 
The new reactor is expected to have a power of 
approximately 3.926 GW and is due to start operations in
March of 2006. In this paper we study the impact 
the flux of $\bar{\nu}_e$
from the new Shika-2 reactor can have 
on the sensitivity of the KamLAND experiment 
to the solar neutrino oscillation parameters.
We present our results for the \kl detector before and after
the starting of the new reactor power plant. 
Finally we  do a global analysis of solar and \kl data
in our analysis including 
the effect of the Shika-2
contribution to the KamLAND signal and
the just published salt data from the
SNO experiment.

%
\section{The $\bar{\nu}_e$ Survival Probability at \kl}
\label{futureprob}
\vspace{-0.3cm}

\begin{figure}[ht]
\begin{center}
\vspace{0.3cm} \epsfxsize = 12cm \epsffile{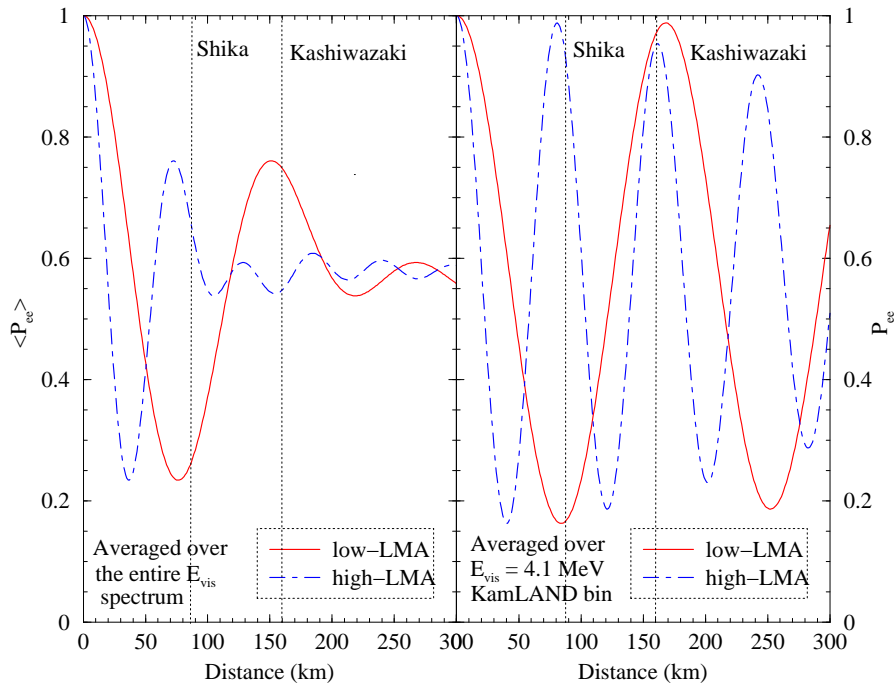}
\leavevmode
\end{center}
\caption{The predicted suppression of the event rate in the \kl
experiment in the cases of low-LMA (solid line) and 
high-LMA (dashed line) solutions, as a 
function of the baseline $L$. The left-hand panel shows the 
probability averaged over the entire positron energy spectrum. 
The right-hand panel shows the survival probability averaged over
the bin with central energy of $E_{vis}=4.1$ MeV and width
of 0.425 MeV. Also marked are 
the distances of the Kashiwazaki and Shika reactor power plants from 
KamLAND.
}
\label{probL}
\end{figure}
\begin{figure}[ht]
\begin{center}
\vspace{0.3cm} \epsfxsize = 12cm \epsffile{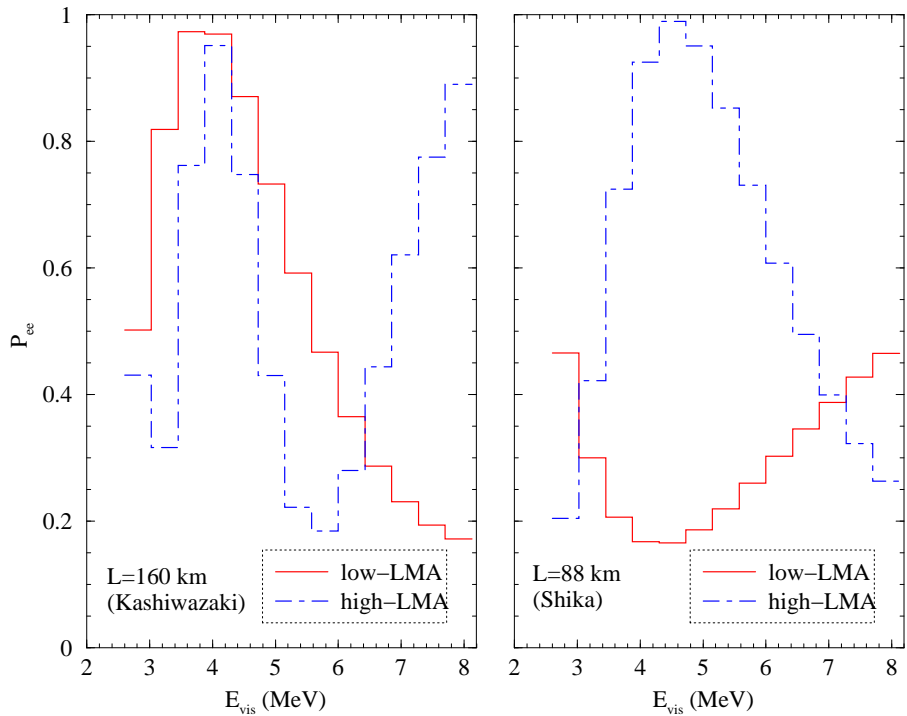}
\leavevmode
\end{center}
\caption{The spectral distortion for the low-LMA (solid line) and 
high-LMA (dashed line) solutions, predicted to be
observed in the \kl experiment if the flux of 
$\bar{\nu}_e$ was produced  by i) the Kashiwazaki reactor 
complex (right-hand panel) and ii) the Shika-2 reactor (left-hand panel).
}
\label{probE}
\end{figure}

   The $\bar{\nu}_e$ flux at \kl detector 
receives contributions 
from a large number of nuclear power plants 
located in different parts of Japan and in
South Korea.
The most powerful is the Kashiwazaki 
reactor complex, which has a total thermal power of 24.6 GW and 
is situated at a distance of about 160 km from KamLAND.
A new reactor power plant, Shika-2, is being built 
a few hundred meters from the site of the existing Shika-1 
reactor, which is located at about 87.7 km from KamLAND.
The thermal power of the new Shika-2 
reactor is about 3.926 GW.  We take the 
distance from Shika-2 to \kl to be 87.7 km. 
It it expected that the Shika-2 plant will
start commercial operation in March 
of 2006. Test operation will start several months
before the commercial operation. 
In this Section we take into account the flux from the new  
Shika-2 reactor and find its impact on the sensitivity 
of KamLAND experiment to the solar neutrino oscillation
parameters $\sin^2\theta_\odot$ and $\ms$.

  The $\bar{\nu}_e$ flux from the various
reactors in Japan (and South Korea) 
are detected in the KamLAND experiment 
through the inverse beta decay 
process $\anue +p \rightarrow e^+ + n$. The 
number of events observed in the \kl detector
is given by 
\be
N_{KL} = N_p \int dE_{vis} \int dE_\nu \sigma(E_\nu) R(E_{vis},E_\nu)
\sum_i \frac{S_i(E_\nu)}{4\pi L_i^2}P_i(\anue \rightarrow \anue)
\label{events}
\ee
%
where $E_{vis}$ is the measured {\it visible} energy of the emitted 
positron, when the true visible energy, $E_{vis}^T 
\cong E_{\nu}-0.80$ MeV,
$E_{\nu}$ being the energy of the incoming $\anue$,
$\sigma(E_\nu)$ is the 
$\anue +p \rightarrow e^+ + n$ reaction
cross-section, 
$S_i(E_\nu)$ denotes the $\anue$ flux 
from the $i$th reactor, 
$L_i$ is the distance between the 
$i$th reactor and \kl, 
$ R(E_{vis},E_\nu)$ is the 
energy resolution function of the detector,
$N_P$ are the number of protons in the target, and
$P_i(\anue \rightarrow \anue)$ is the survival probability of the 
$\anue$ coming from the reactor $i$. 
For the energy resolution we use 
$\sigma(E)/E = 7.5\%/\sqrt{E}$, 
as reported by the \kl collaboration \cite{KamLAND}.
We have used for $N_P$ the number provided 
in \cite{KamLAND}.
The fiducial mass of the detector 
was given as 408 tons.  
Since this may change in the future and the total 
statistics of the experiment depends on 
the mass of detector material 
(in ktons) and the time of exposure (years),
we choose the unit kton-year (kTy) to characterize
the \kl statistics. 

 Apart from the solar neutrino oscillation 
parameters, the $\anue$ survival probability 
$P_i(\anue \rightarrow \anue)$ 
depends on the energy of the antineutrinos and 
the distance $L_i$ between the 
the $i$th reactor and \kl:
\be
P_i(\anue \rightarrow \anue) = 1 - \sin^22\theta_\odot
\sin^2\bigg(\frac{\ms L_i}{4E}\bigg)
\label{prob}
\ee
%
where it is assumed that $\anue$ 
take part in two-neutrino oscillations.
The \kl collaboration presented 
their data in 13 $E_{vis}$ bins 
of width of 0.425 MeV. To get the events in each of 
the \kl bins, 
we integrate \eq{events} between 
$E_{vis} - 0.425/2$ to $E_{vis} + 0.425/2$.

  In Figure \ref{probL} we show the
effects of the survival probability 
$P_i(\anue \rightarrow \anue)$ 
on the \kl event rate as a function of distance. 
The left-hand panel shows the 
probability averaged over 
the entire visible energy spectrum,
while the right-hand panel illustrates
the depletion the \kl detector 
would observe in the bin with central energy $E_{vis}=4.1$ MeV. 
Thus, while the left-hand panel shows 
the predicted suppression of the event rate as 
a function of the reactor distance, the right-hand panel gives 
an idea of the behavior of the observed spectral shape with distance. 
The effects of $P_i(\anue \rightarrow \anue)$
are shown both for the 
now strongly preferred low-LMA ($\ms=7.2\times 10^{-5}$ eV$^2$, 
$\sss =0.3$) and for the high-LMA ($\ms=1.5\times 10^{-4}$ eV$^2$, 
$\sss =0.3$) solutions.
Marked on the figure are also the approximate distances of the 
Kashiwazaki and the Shika reactor power plants from KamLAND.
From the left-hand panel 
we note that the signal 
due to the $\anue$ flux from
Kashiwazaki complex is 
suppressed due to the oscillations by a factor of  
$\sim 0.75~(0.54)$ for the low-LMA (high-LMA) solution. 
The signal due to $\anue$ from the Shika reactor  
is reduced by a factor of $\sim 0.26~(0.64)$ in the 
case of the the low-LMA (high-LMA) solution. The event 
rate observed in \kl with 162 ton-yr of data
corresponds to 0.61 of the rate predicted 
in the absence of oscillations. This is the 
rate observed in the detector taking 
into account the contributions
from all reactors.

As the right-hand panel in Figure \ref{probL} 
indicates, the flux of $\anue$ from the Kashiwazaki complex,
producing $e^{+}$ with  $E_{vis} \sim 4.1$ MeV,
``falls'' in the region of maximum of  
the survival probability (SPMAX),
$P_i(\anue \rightarrow \anue) \cong 1$,
for both the low-LMA and high-LMA solutions. 
In contrast, the $\anue$ with the same energy ($E \sim 3.3$ MeV) 
from the Shika complex located at $\sim 88$ km from KamLAND, 
are affected by a minimum of the survival probability
(SPMIN) in the case of the low-LMA solution, and by a 
maximum (SPMAX) if the true solution is the high-LMA one. 
This feature is further illustrated in Figure \ref{probE},
where we show the positron energy spectrum observed in \kl 
for the two distances of interest -  160 km (Kashiwazaki) and 
88 km (Shika). We have plotted the averaged survival probability 
in each bin for the two solutions.
We note that for $L=160$ km, the spectral shape 
in the statistically important part of the energy spectrum 
is almost the same for both low-LMA and high-LMA solutions, 
with SPMAX at about 4 MeV. Since 
\kl receives most of the
$\anue$ flux from the Kashiwazaki complex, 
this results in the appearance 
of the two allowed solutions: low-LMA and high-LMA. 
For a distance of $\sim $88 km and  
$E_{vis}$ of about 4.5 MeV, however, 
the survival probability of interest exhibits 
a minimum (SPMIN) for the low-LMA solution
and a maximum (SPMAX) for the high-LMA one. 
One could naively expect that a new reactor at the Shika 
site with a large enough thermal power should 
make it easier to discriminate between the currently 
allowed two solutions using the \kl data.
The presence of the SPMIN for 
the $\anue$ flux from Shika in the case of
the low-LMA solution 
might also be expected to
improve the precision of 
the $\sss$ determination from the KamLAND data
\footnote{In \cite{th12} it was argued that a baseline 
of 70 km was ideal for the $\sss$ sensitivity. The Shika distance 
is close to this best baseline.}. 
Since $P_i(\anue \rightarrow \anue)$ has a maximum
in the case of the high-LMA solution, 
one does not expect any improvement in
the precision of the  $\sss$ measurement 
for this solution. 

  Our preceding discussion was concerned with the 
signals in \kl due to the Kashiwazaki or Shika-2 reactors 
alone. However, in a realistic evaluation of the
physics potential of the \kl detector
one has to take into account the contributions 
to the \kl signal from all relevant reactors.

\section{Measurement of the Solar Neutrino 
Oscillation Parameters}
\label{futureconts}
\vspace{-0.3cm}

   In this section we generate 
the prospective \kl data with and 
without the new reactor Shika-2 and study the impact of 
Shika-2 on the determination of 
the solar neutrino oscillation parameters 
using a $\chi^2$ analysis. 
For the errors we assume a
Gaussian distribution and define our $\chi^2$ as 
\bea
\chi^2 =  \sum_{i,j}(N_{i}^{data} - N_{i}^{theory})
(\sigma_{ij}^2)^{-1}(N_{j}^{data} - N_{j}^{theory})~,
\label{chig}
\eea
%
\noindent where $N_i^\alpha$ $(\alpha=data,theory)$ 
is the number of events in 
the $i^{\rm th}$ bin and the sum is over all 
bins. 
We assume that 
\kl will continue to present their data in 13 bins of 
width 0.425 MeV with an energy threshold of 2.6 MeV.
The error correlation matrix $ \sigma_{ij}^2$ contains 
the statistical and systematic errors. While in their first 
paper the \kl collaboration quote a very conservative 6.42\% 
for their systematic error, the latter is expected to 
diminish with time. The most important contribution to the \kl systematic 
error comes from the uncertainty in the knowledge of the fiducial volume
of the detector, which is expected to be reduced. 
The understanding of the other 
systematics would reduce the error further. In our analysis we 
will assume a future \kl plausible systematic error of 5\%. 

\begin{figure}[ht]
\begin{center}
\vspace{0.3cm} \epsfxsize = 12cm \epsffile{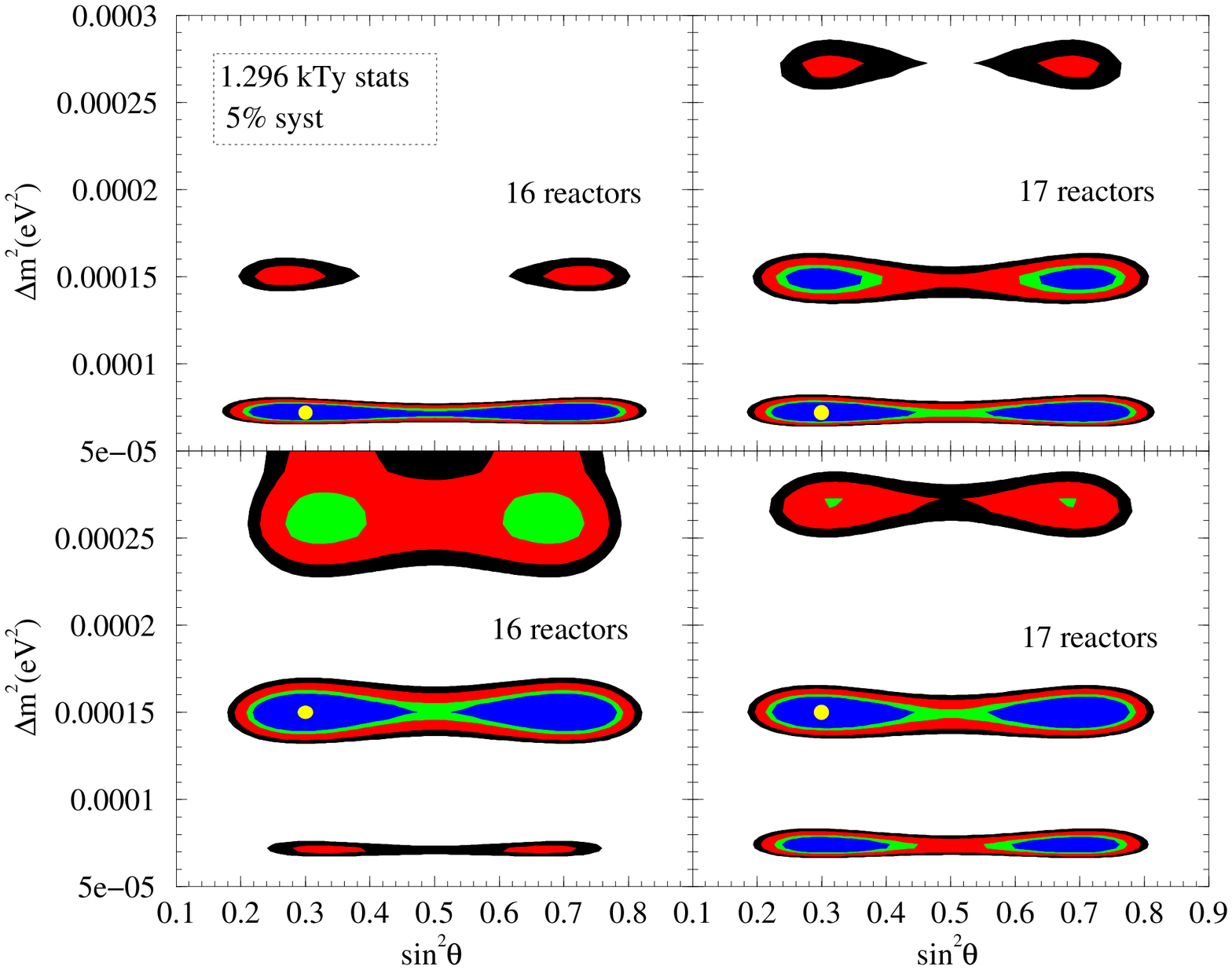}
\leavevmode
\end{center}
\caption{Prospective  90\%, 95\%, 99\% and 99.73\% C.L. 
contours in the $\ms - \sss$ plane,
which would be obtained using the 
\kl data corresponding to 1.296 kTy. In the left-hand 
panels we present results derived with flux from 
the currently operating 16 reactors, giving the 
dominant contribution to the signal in \kl. 
The right-hand panels show the 
corresponding results with the contribution
from the Shika-2 reactor added to that of the indicated 
16 reactors. The upper(lower) panels in both cases give the allowed regions 
when low-LMA (high-LMA) is the correct solution. The points at which 
the spectrum was simulated are shown by yellow circles.
}
\label{cont:1.296}
\end{figure}

\begin{figure}[ht]
\begin{center}
\vspace{0.3cm} \epsfxsize = 12cm \epsffile{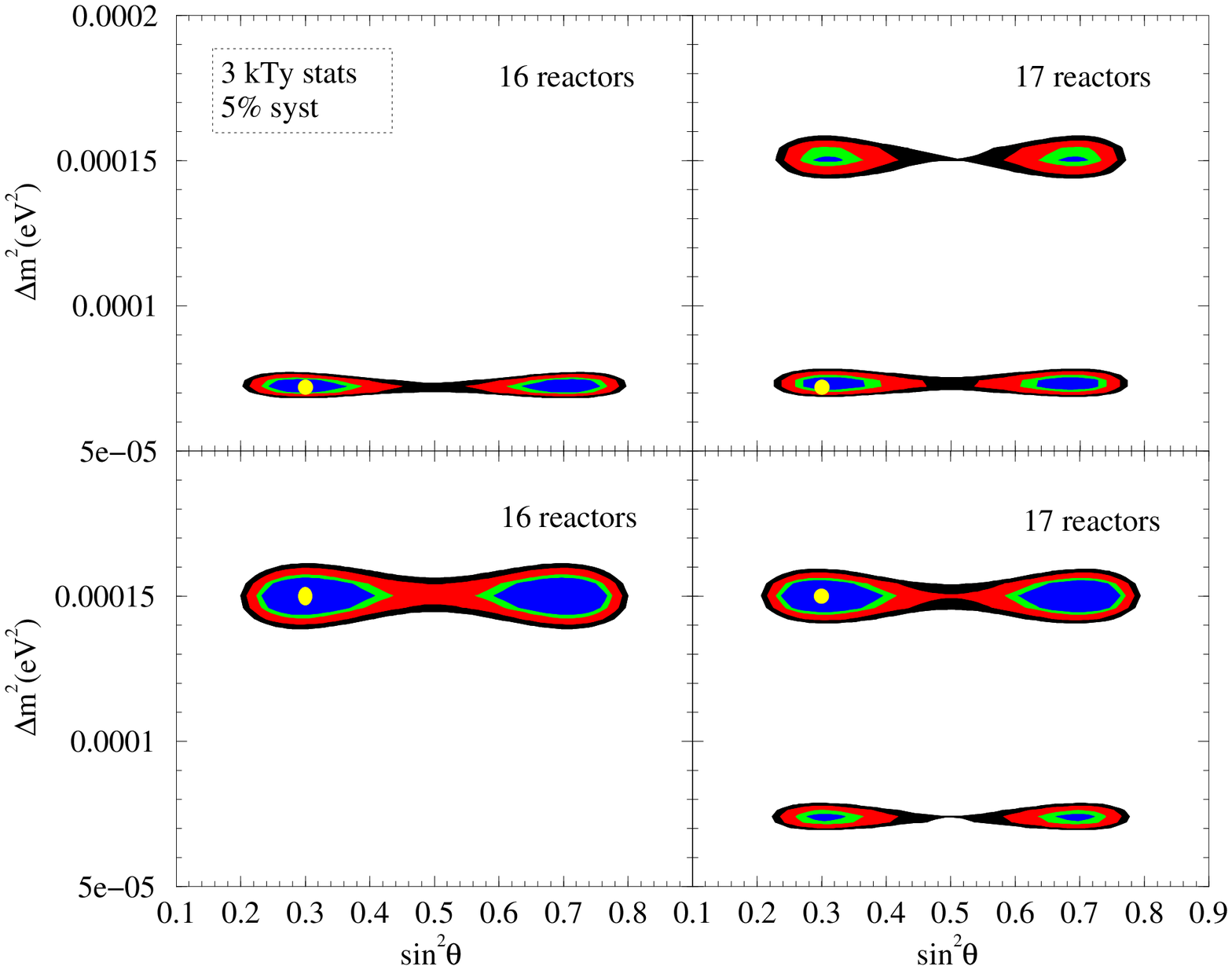}
\leavevmode
\end{center}
\caption{Same as in Figure \ref{cont:1.296}, but for the larger 
statistics of 3 kTy.
}
\label{cont:3}
\end{figure}

  In Figure \ref{cont:1.296} we present the allowed regions in the
$\ms - \sss$ plane, obtained 
when the spectrum measured by \kl is simulated 
at the low-LMA solution best-fit point 
(upper panels) and a high-LMA 
solution point (lower panels).
The left panels show the allowed regions obtained
with the current set-up of reactors in Japan, with 
data foreseen to be collected roughly up to 
March 2006. This corresponds to a statistics of 1.296 kTy. 
The right panels give the corresponding allowed regions
obtained for identical statistics but with the 
contribution from the future Shika-2 reactor 
included. We see that with the current reactor 
set-up in Japan, even the \kl data corresponding to
1.296 kTy would still allow the 
high-LMA (low-LMA) solution at 99\% C.L.
if the low-LMA (high-LMA) is the true solution.
Moreover, if the high-LMA is the correct solution,
there would be additional 
multiple allowed regions at higher values of $\ms$.
Including the contribution from 
the Shika-2 reactor along with that from 
the other reactors, makes the 
ambiguity worse. With the low-LMA (high-LMA) 
as the true solution, 
the spurious high-LMA (low-LMA) regions 
get allowed even at 90\% C.L..
There also appears an additional 
high-LMA solution at $\ms \sim 2.6\times 
10^{-4}$ eV$^2$. Figure \ref{cont:3} shows the corresponding allowed 
regions with KamLAND data of 3 kTy.
While with 3 kTy of data the \kl experiment 
could determine the correct solution
unambiguously with the current set-up of 
contributing reactors, the presence of the 
signal from the Shika-2 reactor  
would make both the low-LMA and high-LMA solutions
allowed at 90\% C.L. and thus would 
not permit to resolve the solution ambiguity.

\begin{figure}[ht]
\begin{center}
\vspace{0.3cm} \epsfxsize = 12cm \epsffile{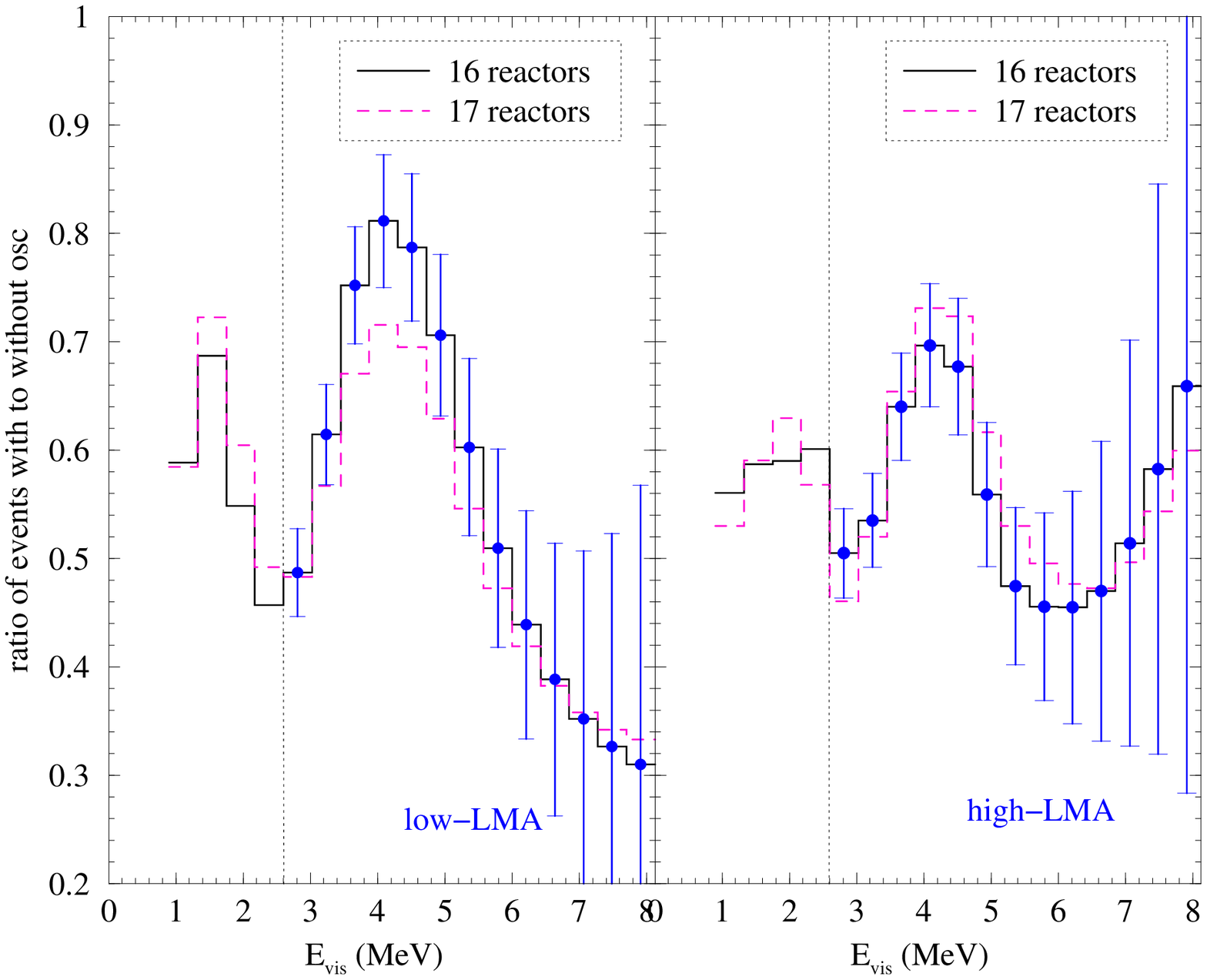}
\leavevmode
\end{center}
\caption{Expected cumulative positron energy spectral distortion in \kl 
for the low-LMA (left-hand panel) and high-LMA (right-hand panel) 
solutions. The solid lines display the distortion expected from 
the present 16 reactor set-up, while the dashed lines give the 
distortion expected with the Shika-2 reactor contribution included.
The error bars show the statistical errors for a KamLAND exposure 
of 3 kTy and with 16 reactor set-up. 
}
\label{prob:total}
\end{figure}

   In Figure \ref{prob:total} we show the ratio of the 
total number of events expected in \kl  
in the case of $\anue$  oscillations 
to the corresponding number predicted in absence 
of oscillations. In this figure we 
take into account
the contributions to the \kl event rate
from all reactors. This is to be compared with Figure \ref{probE} 
which shows the survival probability expected for the 
Kashiwazaki and Shika-2 power plants alone. The left 
panel  of Figure \ref{prob:total} shows the ratios for the case 
with (17 reactors) and without (16 reactors) the contribution of 
Shika-2 reactor, for the spectrum generated 
at the low-LMA solution best fit point. The 
right panel show the corresponding ratio for the 
data generated at the high-LMA solution best-fit point. 

For the low-LMA solution,
we had seen from Figure \ref{probE}
that the signal due to the Kashiwazaki  
$\anue$ flux is affected by a SPMAX, 
while the event rate due to the $\anue$ from
Shika-2 rector 
is affected by a SPMIN. 
The \kl detector measures the power weighted cumulative 
$\anue$ flux from all reactors.
If we consider the effective flux at KamLAND
from the $i^{th}$ reactor plant
without taking into account the $\anue$ oscillations,
\be
\Phi_{KL}^i = \frac{P_i}{4\pi L_i}
\ee
%
we find that for the Kashiwazaki complex 
it corresponds to
$\Phi_{KL}\approx 7.3 \mu W/cm^2$,
while for the Shika-2 reactor 
it is substantially smaller, $\Phi_{KL}\approx 4.1 \mu W/cm^2$.
Thus, the impact on the 
\kl observed signal due to the 
Kashiwazaki $\anue$ flux
will be considerably larger than that
due to the Shika-2 flux,
when the contributions from all the 
reactors are combined.
Hence, for the low-LMA solution,
the result of adding of the Shika-2 signal
on the observed to expected events ratio, is only to 
decrease it somewhat from that obtained without Shika-2.
In other words, even though 
the oscillations generate big spectral 
distortions in the spectra of 
$\anue$ produced at 
Kashiwazaki and at Shika-2 reactors,
which are reflected 
in the corresponding $e^+$ spectra
measured by the \kl experiment, 
since the distortion thus produced
have opposite effects 
(corresponding to SPMAX and SPMIN respectively), 
the net spectral distortion 
observed in \kl would be
smaller than in the case of no Shika-2
contribution.
 This results in increased ambiguity in the determination
of $\ms$, which depends strongly on the magnitude of the distortion 
in the measured resultant spectrum. This can be seen in the left 
panel of Figure \ref{prob:total}. 
For the high-LMA solution, both Kashiwazaki and Shika-2 
contributions to the spectrum measured at \kl
are affected by a SPMAX. This results in 
a mild increase of the spectral distortion 
for the high-LMA solution with the introduction 
of Shika-2 signal. Most importantly,
as this Figure indicates, after turning on the Shika-2 reactor, 
the differences between the predicted spectral shapes 
in the cases of the low-LMA and 
the high-LMA solution diminish. This explains the increase in the 
degeneracy between the two solutions, as 
observed in Figure \ref{cont:1.296}. This degeneracy 
is not lifted even when statistics is increased to 
3 kTy, as seen in Figure \ref{cont:3}.

\begin{figure}[ht]
\begin{center}
\vspace{0.3cm} \epsfxsize = 12cm \epsffile{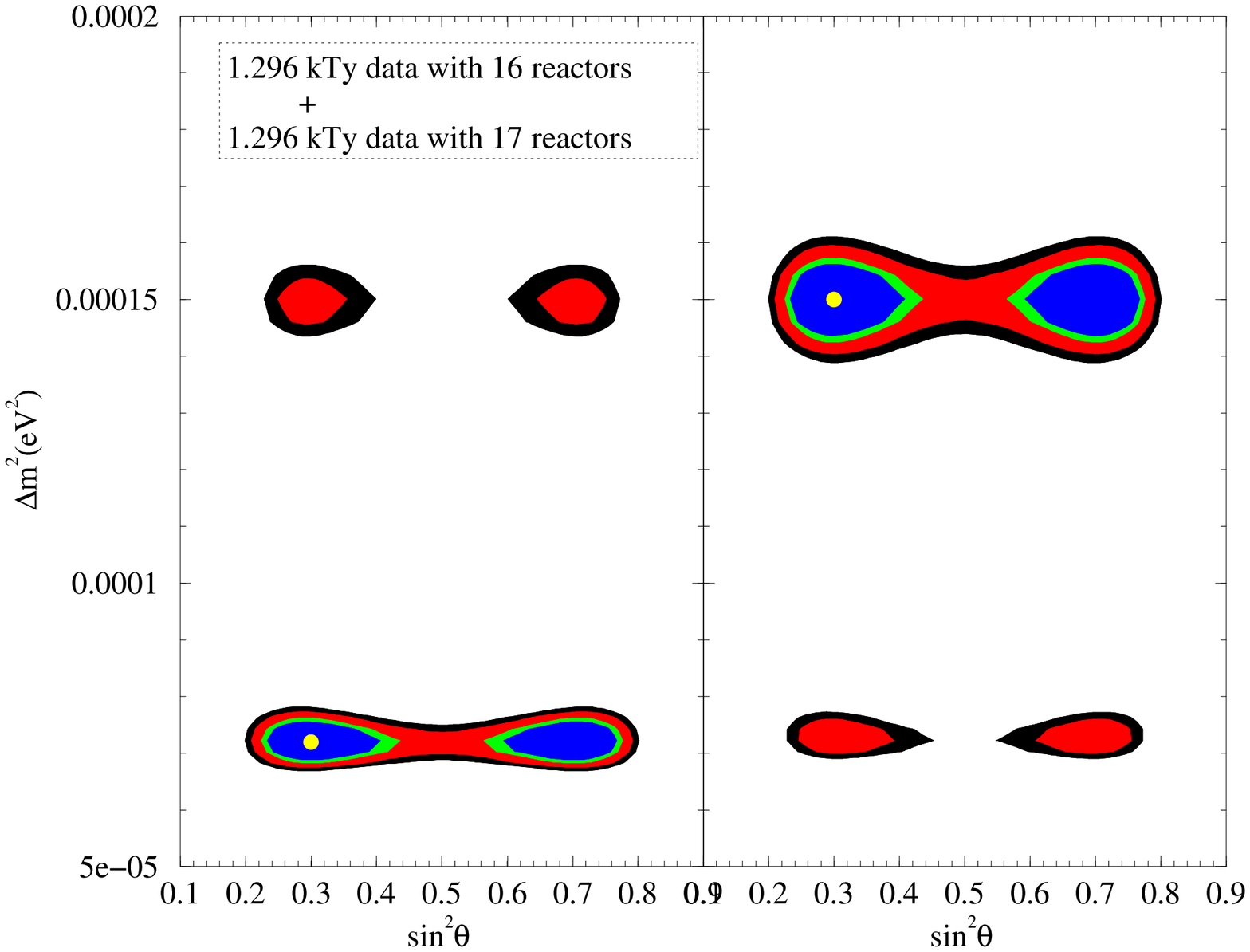}
\leavevmode
\end{center}
\caption{The C.L. contours obtained from a combined analysis of 
1.296 kTy of data collected with 16 reactors as $\bar{\nu}_e$ sources, and 
an additional 1.296 kTy of data collected when 17 reactors 
including Shika-2 contribute to the signal in the \kl experiment.
}
\label{cont:16+17}
\end{figure}
\begin{table}
\begin{ruledtabular}
\begin{tabular}{cccc}
Reactor & true & Statistics & 99\% C.L. range \cr
set-up & solution &  & for $\sss$ \cr
\hline
16 & low-LMA & 1.296 kTy & $0.18-0.81$   \cr
 16 & high-LMA & 1.296 kTy & $0.19-0.81$    \cr
 17 & low-LMA & 1.296 kTy & $0.19-0.80$    \cr
 17 & high-LMA & 1.296kTy  & $0.19-0.80$    \cr
 16 & low-LMA & 3.0 kTy & 0.$21-0.45 $   \cr
 16 & high-LMA & 3.0 kTy & 0.$21-0.79$    \cr
 17 & low-LMA & 3.0 kTy  & $0.24-0.46$    \cr
 17 & high-LMA & 3.0 kTy  & $0.21-0.78$    \cr
 16+17 & low-LMA & 1.296 kTy + 1.296 kTy & $0.21-0.79$    \cr
\end{tabular}
\caption{\label{tab1}
The 99\% C.L. allowed range of $\sss$ expected from future data.
The first column gives the number of main reactors 
acting as $\anue$ source for KamLAND. The final column gives the 
99\% C.L. range of allowed values of $\sss$ for the different 
data sets.
}
\end{ruledtabular}
\end{table}

   Up to now we have been discussing a scenario where we either have 
16 reactors without Shika-2 or we have 17 reactors with Shika-2.
The more realistic scenario will be the one where 
\kl collects data with the present 16 reactors until Shika-2 starts 
operation and thereafter \kl will observe the flux from all 17 reactors.
It is expected that Shika-2 would begin to operate in March 2006. 
This implies that \kl would have already 
collected 4 years of data corresponding to 
1.296 kTy statistics with the current fiducial volume. In 
Figure \ref{cont:16+17} we present the allowed regions of values of
$\ms$ and $\sss$ 
in this more realistic scenario. 
We consider an aggregate \kl 
statistics of 1.296 kTy with the current 16 reactors and
a further 1.296 kTy data with 17 reactors including Shika-2.
We assume the same systematic uncertainty of 5\% for both the 
data sets. The Figure shows that with such a 
combined data set the 
ambiguity between the low-LMA and high-LMA solutions 
appears only at 
99\% C.L., as in the case with just the 16 
reactors.
It should be noted, however,
that even with a total of $2\times 1.296\approx 2.6$ 
kTy data, \kl still would not be able to
resolve the ambiguity between the two solutions. 
This should be compared with the 
3 kTy results for the 16 reactor set-up given in Figure 
\ref{cont:3}, which show that the ambiguity 
between the two solutions could be resolved even at the 
$3\sigma$ level. Thus, contrary to what one might naively expect, the 
advent on Shika-2 will reduce the  
sensitivity of KamLAND to $\ms$. 

  Next we focus on the impact of the new Shika-2 reactor on 
the precision of the measurement of $\sss$. 
From the Figures \ref{cont:1.296}, 
\ref{cont:3} and \ref{cont:16+17} we find that there is hardly 
any improvement in the precision of $\sss$ determination
with the inclusion 
of Shika-2 contribution, contrary to what might be 
expected.
In Table \ref{tab1} we present the 99\% C.L. allowed 
ranges of values of 
$\sss$. The allowed range of $\sss$
for the low-LMA solution 
diminishes only marginally with the inclusion of the Shika-2 
contribution, owing to the effect of the SPMIN, as discussed 
in \cite{th12}. Only the case with 17 reactors and 3 kTy statistics 
shows some improvement over the 16 reactor case.
The improvement here is much less than that obtained in 
\cite{th12} for two major reasons. Firstly, Shika-2 has a thermal power 
of only 3.926 GW while the reactor considered in \cite{th12} had 
power of 24.3 GW, a la Kashiwazaki. Secondly, in \cite{th12} the 
effect of the presence of other more powerful reactors apart from the 
reactor producing the SPMIN, was not taken into account. 
The presence of the other powerful reactors at different baselines 
(Kashiwazaki, in particular)
changes the shape of the resultant positron spectrum at \kl from one 
corresponding to a SPMIN to one corresponding to a SPMAX, as can be seen 
in Figure \ref{prob:total}.

\begin{figure}[ht]
\begin{center}
\vspace{0.3cm} \epsfxsize = 12cm \epsffile{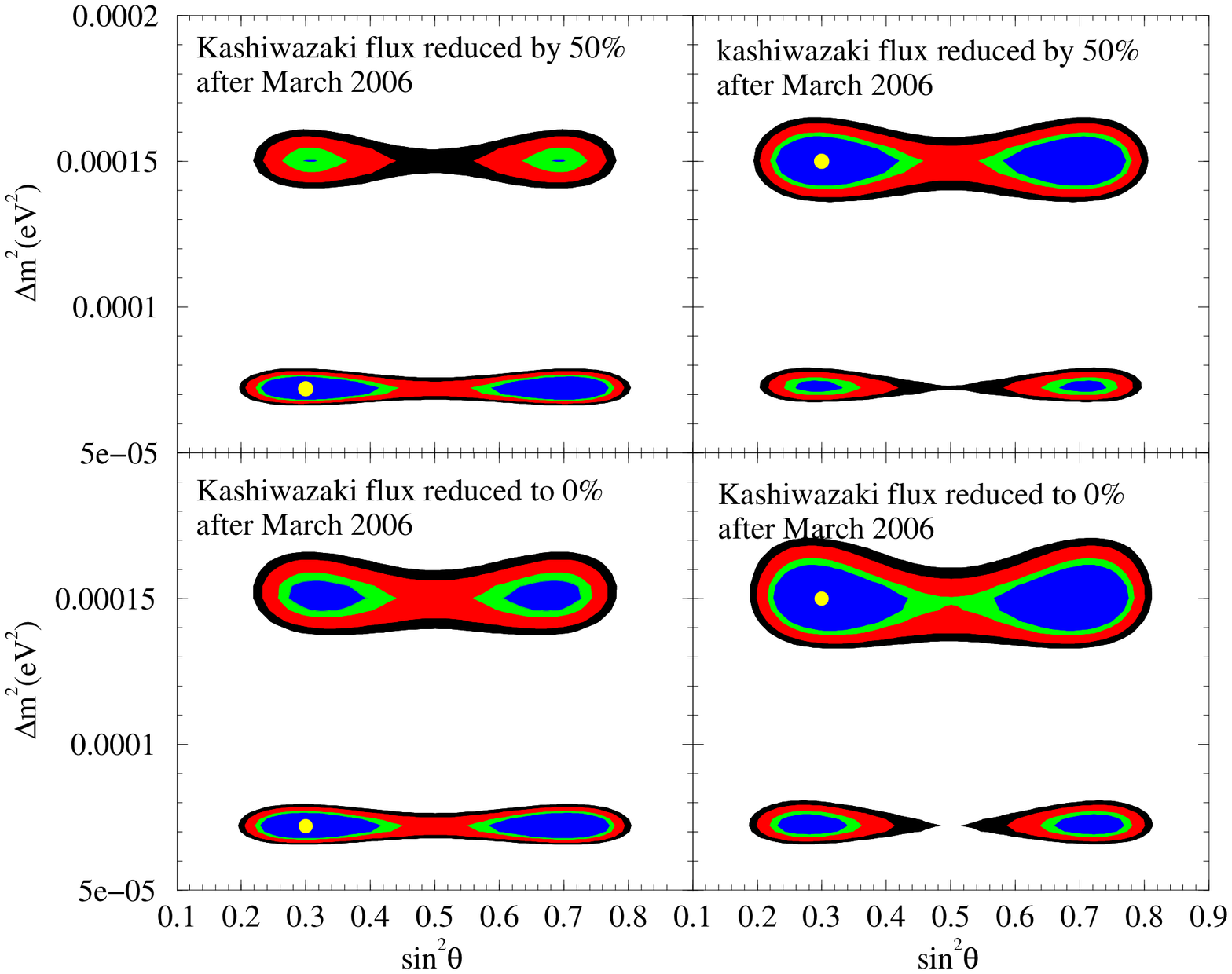}
\leavevmode
\end{center}
\caption{Same as in Figure \ref{cont:16+17}, but with the 
Kashiwazaki power reduced by 50\% (upper panels) and 
kashiwazaki power switched off completely (lower panels) after March 
2006, when Shika-2 reactor would start operation.
}
\label{cont:kashi0}
\end{figure}
\begin{figure}[ht]
\begin{center}
\vspace{0.3cm} \epsfxsize = 12cm \epsffile{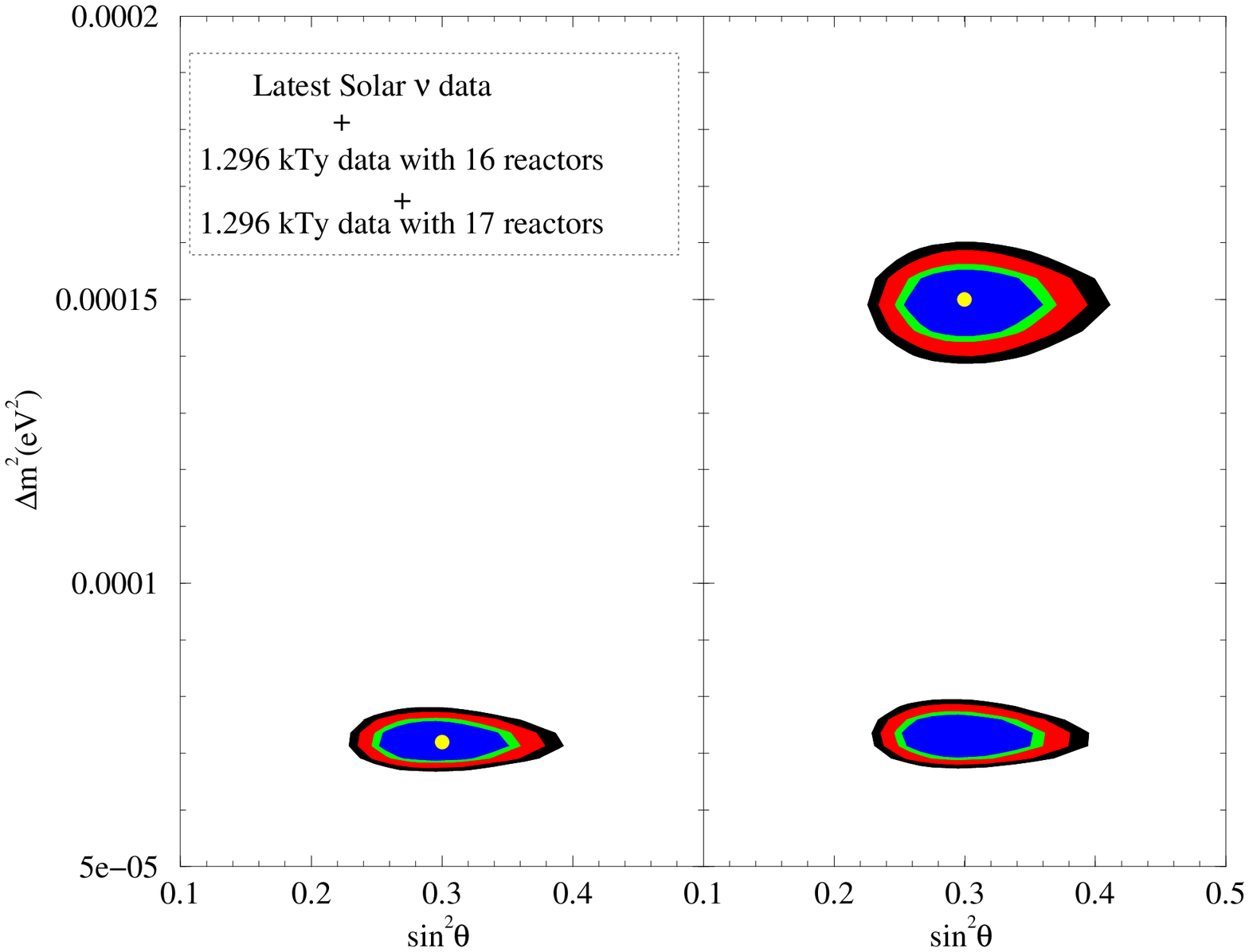}
\leavevmode
\end{center}
\caption{The combined allowed C.L. contours expected after combining 
the global solar neutrino data, including the salt enriched SNO results, 
with the total data obtained 
from \kl after 1.296 kTy data with 16 reactors and 1.296 kTy 
data with the 17 reactors including Shika-2. The left-hand 
panel corresponds to low-LMA as true solution while the 
right-hand panel gives the allowed areas obtained if 
high-LMA was true. 
}
\label{cont:global}
\end{figure}

We consider next the hypothetical case of 
the Kashiwazaki complex operating with reduced power, which would make
the SPMIN associated with the Shika-2 
contribution more pronounced.
The corresponding results are presented 
in Figure \ref{cont:kashi0}.
This figure is obtained 
supposing that \kl receives flux from the current 16 reactors, including 
the Kashiwazaki reactors running at full power up to March 2006
and that thereafter it  
receives flux from Shika-2 power plant as well for another period of 
4 years, giving a total statistics of $1.296+1.296$ kTy for the 
two phases combined. For the second phase, however,
we assume that the power of the 
Kashiwazaki reactor complex is reduced 
by (i) 50\%, and (ii) 100\%, i.e., 
a total shut-down of the Kashiwazaki reactors. The 
impact on the determination of the solar neutrino oscillation 
parameters in  the former case is shown in 
the upper panels of Figure \ref{cont:kashi0},
while the results for the latter case are 
displayed in the lower panels of Figure \ref{cont:kashi0}. 
We note that in neither case
one can discriminate
between the low- and high- LMA solutions
and/or can obtain better constraints on  
$\sss$. In fact, the precision gets worse, owing mainly to the fact 
that the statistics of the \kl experiment decreases significantly 
if the $\bar{\nu}_e$ flux from the Kashiwazaki complex is cut down.

Finally, we present the constraints obtained 
on $\ms$ and $\sss$ when the data from the 
\kl experiment is combined with the global solar neutrino 
data. In Figure \ref{cont:global} we show the 
allowed regions 
derived 
by combining the global solar neutrino data (including the 
latest salt phase data from SNO) with the 
combined 1.296 kTy \kl data  
collected with 16 reactors and 1.296 kTy \kl 
data obtained with 17 reactors, 
including the Shika-2 flux. A comparison of 
Figure \ref{cont:16+17} with Figure \ref{cont:global} shows
that the inclusion of the solar neutrino data rules out the 
spurious high-LMA solution if low-LMA is the correct one. However, 
if the positron spectrum measured at \kl would correspond to a 
$\ms$ in the high-LMA region,
the solar neutrino data would
further increase the degeneracy and make the 
low-LMA 
solution allowed 
at 90\% C.L.. In fact, the best-fit $\ms$ and $\sss$ 
for the combined analysis come in the low-LMA zone.
The reason for 
this behavior can be traced to the fact that the solar neutrino
data in general, and the SNO data in particular, 
favor the low-LMA solution over the high-LMA one.

\section{Conclusions}
\label{section:conclusions}
\vspace{-0.3cm}

In the present article we have studied the impact 
of the $\anue$ flux from a new 
reactor, Shika-2, 
on the sensitivity of the KamLAND experiment 
to the solar neutrino oscillation parameters,
$\ms$ and $\sss$. This upcoming reactor is 
proposed to have a thermal power of
3.926 GW  and would be located at a distance of about 88 km 
from the \kl detector.
Since the distance of 88 km corresponds to a minimum (SPMIN) in the 
resultant positron spectrum at \kl,
it was expected that the introduction of this detector may improve
the precision with which the mixing angle $\sss$ can be determined. 
However, it follows from our study that the
precision of the measurement of
the solar neutrino mixing angle, $\theta_{\odot}$,
is unlikely to be improved by  the addition of this new 
reactor
because of the averaging effect of fluxes 
from the other reactors. 
In fact, once the contribution of Shika-2 
to the \kl event rate is taken into account, 
the precision to $\ms$ from the \kl experiment
worsens, 
increasing the ambiguity between the low-LMA and high-LMA solutions. 
For the most realistic case where we include the effect of Shika-2 after 
March 2006 and consider about 2.6 kTy statistics in the simulated \kl 
spectrum, we find that 
the degeneracy between the 
two solutions remain at 99\% C.L. 
from the \kl data alone,
irrespective of whether the \kl spectral data is simulated 
in the low-LMA or high-LMA solution region.  
We have also  presented results on
prospective determination of  
$\ms$ and $\sss$ from the combined \kl and the solar neutrino data, including  
the just published salt phase data from the 
SNO experiment. 
Since the SNO salt data disfavours the higher $\ms$ regions,
the high-LMA solution is found to be absent
if the \kl spectrum is simulated in the low-LMA zone. 
However, if the observed \kl spectrum corresponds to
oscillation parameters in the high-LMA zone, 
the ambiguity 
between the low-LMA and high-LMA solutions would remain
as a result of the conflicting trends of the \kl and the 
current solar neutrino data.


\vskip 1 cm
\leftline{\bf Acknowledgements.} 
We would like to thank F. Suekane for useful 
correspondence. S.T.P. acknowledges informative
discussions with K. Heeger.
This work was supported in part 
by the Italian MIUR and INFN under the programs 
``Fenomenologia delle Interazioni Fondamentali'' 
 (S.T.P. and S.G.) and ``Fisica Astroparticellare'' (S.C.).

\vspace{-0.3cm}


\begin{thebibliography}{50}


\bibitem{Cl98}   B.T. Cleveland {\em et al.}, 
                {\em  Astrophys. J.} {\bf 496} (1998) 505;
                Y.\ Fukuda {\em et al.},
               {\em  Phys.\ Rev.\ Lett.\ } {\bf 77} (1996) 1683;
                V.\ Gavrin, {\em  Nucl. Phys. Proc. Suppl.} {\bf 91} (2001) 36;
                W.\ Hampel {\em et al.},
               {\em  Phys.\ Lett.\ } {\bf B447} (1999) 127;
                M.\ Altmann {\em et al.},
               {\em  Phys.\ Lett.\ } {\bf B490} (2000) 16.

\bibitem{SKsol} Super-Kamiokande Coll.,
                Y. Fukuda {\em et al.}, 
              {\em  Phys.\ Rev.\ Lett. } {\bf 86} (2001) 5656 and 5651.
%
%

\bibitem{SNO1} SNO Coll.,
               Q.R. Ahmad \textit{et al.}, 
{\em Phys. Rev. Lett.} {\bf 87} (2001) 071301.

\bibitem{SNO2} SNO Coll.,
               Q.R. Ahmad \textit{et al.}, 
{\em Phys. Rev. Lett.} {\bf 89} (2002) 011302 and 011301. 


\bibitem{KamLAND} KamLAND Coll., K. Eguchi {\em et al.}, 
{\em Phys.\ Rev.\ Lett.}  {\bf 90} (2003) 021802.

\bibitem{STPSchlad97} S.T. Petcov, Lecture Notes in Physics, 
v. {\bf 512} (eds. H. Gausterer and C.B. Lang, Springer, 1998), p. 281
(arXiv:hep-ph/9806466); S.~Goswami,Pramana {\bf 60}, 261 (2003),
(arXiv:hep-ph/0305111);
 M.C. Gonzalez-Garcia and Y. Nir,
               {\em Rev.\ Mod.\ Phys.}  {\bf 75} (2003) 345. 

\bibitem{RSFP} C.-S. Lim and W. Marciano, 
{\em Phys. Rev.} {\bf D37} (1988) 1368;
E.Kh. Akhmedov, {\em Phys. Lett.} {\bf B213} (1988) 64.

\bibitem{FCNC} M.M. Guzzo, A. Masiero and S.T. Petcov, 
{\em Phys. Lett. } {\bf B260} (1991) 154;
E. Roulet, {\em Phys. Rev.} {\bf D44} (1991) R935.

\bibitem{VWEPVLI} M. Gasperini, {\em Phys. Rev.} {\bf D39} (1989) 3606;
A. Halprin and C.N. Leung, {\em Phys.\ Rev.\ Lett.}  {\bf 67} (1991) 1833;
S. Coleman and S. Glashow, {\em Phys. Lett. } {\bf B405} (1997) 249.

\bibitem{Pont4667} B. Pontecorvo, Chalk River Lab. report PD--205, 1946;
{\em Zh. Eksp. Teor. Fiz.} {\bf 53} (1967) 1717.

\bibitem{Davis68} R. Davis, D.S. Harmer and K.C. Hoffman,
Phys. Rev. Lett. {\bf 20}, 1205 (1968); Acta Physica Acad.
Sci. Hung. {\bf 29} Suppl. 4, 371 (1970); R. Davis,
Proc. of the ``Neutrino~`72''~Int. Conference, Balatonfured,
    Hungary, June 1972 (eds. A. Frenkel and G.  Marx,  OMKDK-TECHNOINFORM,
    Budapest, 1972), p. 5.


%
%
%
%
%
%
%
%
%
%
%
%
%
%
%
%
%

\bibitem{solfit1}
A.~Bandyopadhyay, 
S.~Choubey, R.~Gandhi, S.~Goswami and D.~P.~Roy,
Phys.\ Lett.\ B {\bf 559}, 121 (2003) [arXiv:hep-ph/0212146];

\bibitem{solfit2} G.~L.~Fogli et al., 
Phys.\ Rev.\ D {\bf 67}, 073002 (2003)
[arXiv:hep-ph/0212127];
%
M.~Maltoni, T.~Schwetz and J.~W.~Valle,
arXiv:hep-ph/0212129;
%
J.~N.~Bahcall, M.~C.~Gonzalez-Garcia and C.~Pena-Garay,
JHEP {\bf 0302}, 009 (2003) [arXiv:hep-ph/0212147];
%
H.~Nunokawa, W.~J.~Teves and R.~Zukanovich Funchal,
arXiv:hep-ph/0212202;
%
P.~Aliani et al., 
arXiv:hep-ph/0212212;
%
P.~C.~de Holanda and A.~Y.~Smirnov,
JCAP {\bf 0302}, 001 (2003) [arXiv:hep-ph/0212270].
%
%

\bibitem{SNO3} SNO Coll., S.N. Ahmed et al., arXiv:nucl-ex/0309004.

\bibitem{Maris:2002cv}
M.~Maris and S.~T.~Petcov,
Phys.\ Lett.\ B {\bf 534}, 17 (2002)
[arXiv:hep-ph/0201087].

\bibitem{SNO3ADSSS} 
A.~Bandyopadhyay, 
S.~Choubey, S.~Goswami, S.T. Petcov and D.~P.~Roy,
hep-ph/0309174. 

\bibitem{others}
M.~Maltoni, T.~Schwetz, M.~A.~Tortola and J.~W.~Valle,
arXiv:hep-ph/0309130.
P.~Aliani, V.~Antonelli, M.~Picariello and E.~Torr
ente-Lujan,
arXiv:hep-ph/0309156.



\bibitem{SKatm9802}
Y.~Fukuda {\it et al.}  [Super-Kamiokande Collaboration],
Phys.\ Rev.\ Lett.\  {\bf 81}, 1562 (1998) [arXiv:hep-ex/9807003];
                 M.~Shiozawa,
                 talk given at the Int. Conf. on Neutrino Physics and 
                 Astrophysics ``Neutrino'02'', May 25 - 30, 2002, 
                 Munich, Germany.


\bibitem{SKatmo03} Super-Kamiokande Coll., 
Y. Hayato \textit{et al.}, Talk given at the Int. EPS Conference
on High Energy Physics, July 17 - 23, 2003, Aachen, Germany.

\bibitem{SPNDM03} S.T. Petcov, Invited talk given at the First Yamada
Symposium on Neutrinos and Dark Matter in Nuclear Physics,
June 9 - 14, 2003, Nara, Japan, Ref. SISSA 72/03/EP
(http://ndm03.phys.sci.osaka-u.ac.jp).

\bibitem{Maris:2000DN}
M.~Maris and S.~T.~Petcov,
Phys.\ Rev.\ D {\bf 62}, 093006 (2000)
[arXiv:hep-ph/0003301].

%
%

\bibitem{prekl}
A.~Bandyopadhyay, S.~Choubey, R.~Gandhi, S.~Goswami and D.~P.~Roy,
arXiv:hep-ph/0211266.

\bibitem{preklall}
V.~D.~Barger, D.~Marfatia and B.~P.~Wood,
Phys.\ Lett.\ B {\bf 498}, 53 (2001)
[arXiv:hep-ph/0011251];
%
H.~Murayama and A.~Pierce,
Phys.\ Rev.\ D {\bf 65}, 013012 (2002)
[arXiv:hep-ph/0012075];
%
A.~de Gouvea and C.~Pena-Garay,
Phys.\ Rev.\ D {\bf 64}, 113011 (2001)
[arXiv:hep-ph/0107186].


\bibitem{th12}
A.~Bandyopadhyay, S.~Choubey and S.~Goswami,
Phys.\ Rev.\ D {\bf 67}, 113011 (2003)
[arXiv:hep-ph/0302243].

\bibitem{Bahcall:2003ce}
J.~N.~Bahcall and C.~Pena-Garay,
arXiv:hep-ph/0305159.

\bibitem{Petcov:2001sy}
S.~T.~Petcov and M.~Piai,
Phys.\ Lett.\ B {\bf 533}, 94 (2002)
[arXiv:hep-ph/0112074].


\bibitem{th12hlma}
S.~Choubey, S.~T.~Petcov and M.~Piai,
arXiv:hep-ph/0306017.

\bibitem{shika2}
K.~Nakamura, Talk at the 5th International Workshop on Neutrino Factories 
\& Superbeams (NuFact'03), Columbia University, New York, June 5-11, 2003;
{\it http://www.cap.bnl.gov/nufact03/}

\end{thebibliography}
\end{document}